\pdfoutput=1
\documentclass{JINST}

\usepackage{amsmath}

\title{Model selection for amplitude analysis}
\author{Baptiste Guegan, John Hardin, Justin Stevens and Mike Williams \\
Massachusetts Institute of Technology, Cambridge, MA, United States
}

\abstract{ 

Model complexity in amplitude analyses is often {\em a priori} under-constrained since the underlying theory permits a large number of possible amplitudes to contribute to most physical processes.  
The use of an overly complex model results in reduced predictive power and worse resolution on unknown parameters of interest.
Therefore, it is common to reduce the complexity by removing from consideration some subset of the allowed amplitudes.
This paper studies a method for limiting model complexity from the data sample itself through regularization during regression in the context of 
a multivariate (Dalitz-plot) analysis.  
The regularization technique applied greatly improves the performance.  
An outline of how to obtain the significance of a resonance in a multivariate amplitude analysis is also provided.
 }

\begin{document}

%%%%%%%%%
%  Introduction  %
%%%%%%%%%
\section{Introduction}
\label{Sec:Intro}

Model complexity in amplitude analyses is often {\em a priori} under-constrained since the underlying theory permits a large number of possible amplitudes to contribute to most physical processes.  
The use of an overly complex model results in reduced predictive power and worse resolution on unknown parameters of interest, {\em e.g.}, resonance parameters or the contribution from a particular amplitude.  
Therefore, it is common to reduce the complexity by removing from consideration some subset of the allowed amplitudes.
For example, often times only the lowest few possible values of orbital angular momenta are considered.
In some cases {\em ad hoc} procedures are applied in an attempt to make this reduction in complexity unbiased.  Such procedures include starting from a simple model and considering one additional amplitude at a time and labeling the amplitude as "unimportant" if the $\chi^2$ or likelihood does not change by a pre-defined "large enough" amount.  
This {\em stepwise} model reduction is well known to not be optimal~\cite{Harrell}.
Furthermore, it is not well suited for amplitude analysis due to the fact that the most important contribution from an amplitude may be due to its interference with another amplitude.  The stepwise procedure risks never considering two interfering amplitudes in the same model.  
This paper considers alternatives for selecting from data a set of amplitudes to use in the fit model in the context of a Dalitz-plot analysis.

\section{Example analysis}
\label{Sec:2D}

The model we consider in this paper is production of an $abc$ final state, where $a$, $b$ and $c$ are pseudoscalars (\textit{i.e.}, they have spin-parity $0^{-}$) with masses $m_a = m_b = m_c = 0.1$ in arbitrary units.  This final state will first be studied in bins of $abc$ invariant mass, $m_{abc}$, with the fits performed in each bin being independent.  The particles that decay into the $abc$ final state are generically labeled as $X$, {\em e.g.}, the reaction might be $\gamma p \to pX, X\to abc$.
The decay amplitudes are constructed in the isobar formalism with the observed intensity written as the coherent sum of resonant terms plus a non-resonant term added incoherently as follows:
\begin{equation}
|\mathcal{M}(\vec{x},m_{abc})|^2 = \sum_{M} \left| \sum_k a_k e^{i \phi_k} \mathcal{A}^M_k (\vec{x}) \right|^2 + |a_{nr}|^2,
\label{Eqn:Model}
\end{equation}
where $\vec{x} = (m_{ab}^2, m_{ac}^2)$ represents the position in the Dalitz plot, $a_k e^{i \phi_k}$ describes the unknown complex factor for each resonant component of the model, denoted by the index $k$, and $M$ denotes the spin substates of the various $X$ particles\footnote{We drop the explicit $m_{abc}$ dependence on the right side of Eq.~\ref{Eqn:Model}.  All quantities, including the complex constants, depend on $m_{abc}$.}.
Table~\ref{Table:ModelParams} lists the properties and decay channels of the resonances contained in the true probability density function (p.d.f.) for the primary model considered in this paper, referred to as Model I.
The resonant terms contain contributions from Blatt-Weisskopf barrier factors~\cite{Blatt}, relativistic Breit-Wigner line shapes to describe the propagators, and spin factors obtained using the Zemach formalism~\cite{Zemach}.  The amplitudes are evaluated using the \texttt{qft++} package~\cite{Williams:2008wu}.  

The key to this type of analysis is the fact that the Dalitz-plot distributions can be used to separate out the various partial-wave contributions.  This can be done in each $m_{abc}$ bin without the need to assume some resonant-like behavior.  A second step in such an analysis is to fit the magnitudes and phase differences observed {\em vs} $m_{abc}$ to determine if resonances do contribute and, if so, extract values for their pole parameters.    

%%%% Table %%%%
\begin{table}[]
  \centering
  \begin{tabular}{|c|c|c|c|c|c|c|c|c|}
    \hline
    $J^P$ & $m_X$ & $\Gamma_X$ & $M$ & $j$ & $l$ & Isobar mass & Isobar width & Isobar daughters \\
    \hline
    $1^+$ & 1.00 & 0.30 & 0 & 1 & 0 & 0.75 & 0.10 & $bc$ \\
    $2^-$ & 1.45 & 0.25 & 0 & 2 & 0 & 1.10 & 0.15 & $bc$ \\
    $2^-$ & 1.45 & 0.25 & 0 & 1 & 1 & 0.60 & 0.10 & $ab,ac$ \\
    $2^+$ & 1.10 & 0.15 & 1 & 1 & 2 & 0.60 & 0.10 & $ab,ac$ \\
    $1^+$ & 1.25 & 0.25 & 1 & 1 & 0 & 0.75 & 0.10 & $bc$ \\
    non-resonant & & & & & & & & $abc$ \\
    \hline
  \end{tabular}
  \caption{ The set of resonant terms and properties used in the true p.d.f.\ of Model~I for the Dalitz-plot example.  The symbols are: $J^P$, $m_X$, $\Gamma_X$, $M$ the spin-parity, mass, width and spin projection of $X$;  $j$ is the isobar spin; and $l$ is the orbital angular momentum between the isobar and the bachelor particle.}
  \label{Table:ModelParams}
\end{table}

%%%% Figure %%%%
\begin{figure}
        \begin{center}
                \includegraphics[width=1.0\textwidth]{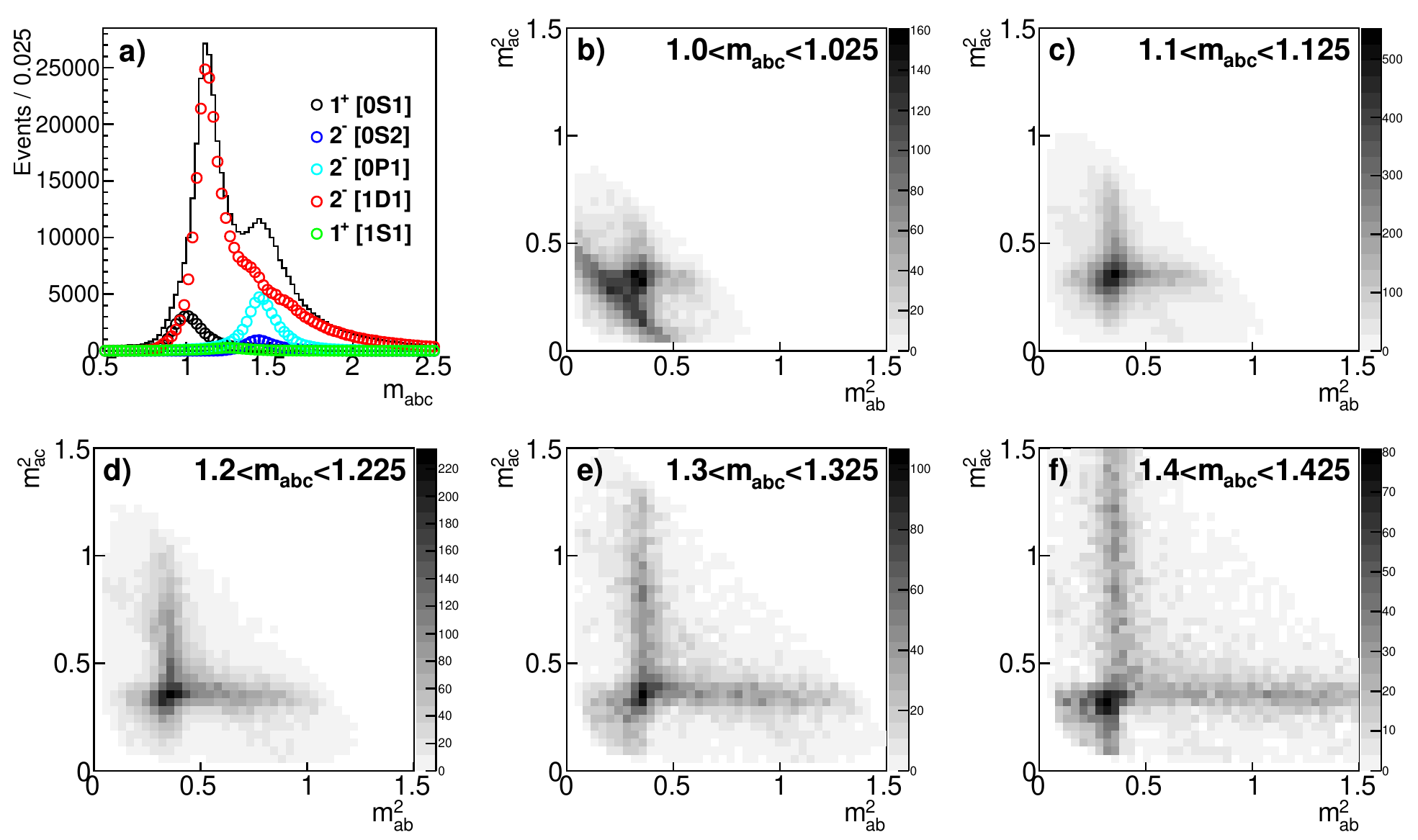}
                \caption{ a) Invariant mass distribution of the $abc$ system for one simulated data set overlaid with the contributions from the individual resonant amplitudes.  b) - f) Dalitz plot distributions in a few bins of $m_{abc}$.}
                \label{fig:DalitzMass}
        \end{center}
\end{figure}

Figure~\ref{fig:DalitzMass} a) shows the invariant mass distribution of the $abc$ system for a simulated data set containing $\sim4.4\times10^5$ events.  There is clear structure from the resonant amplitudes in the model.  The Dalitz plot distributions for a few $m_{abc}$ bins of Fig.~\ref{fig:DalitzMass} a) are shown in panels b) - f); these demonstrate the variation of the resonant structure with $m_{abc}$ in the generated model.
The goal of the amplitude analysis in this example is to determine the complex parameters, $a_k e^{i \phi_k}$, from an extended maximum likelihood fit to the data performed in 0.025-wide bins of $m_{abc}$.  This so-called mass-independent fit is typical in hadron spectroscopy, where it is desirable to avoid parameterization of the produced $X$ resonances.  The mass dependence of the $a_k e^{i \phi_k}$ factors is then studied separately to search for new resonances or to measure properties of known ones.
An ensemble of simulated data sets that contain $500-25\,000$ events per $m_{abc}$ bin is used in this study, where the typical number of events in a given $m_{abc}$ bin is shown in Figure~\ref{fig:DalitzMass} a).

\begin{figure}
        \begin{center}
                \includegraphics[width=1.0\textwidth]{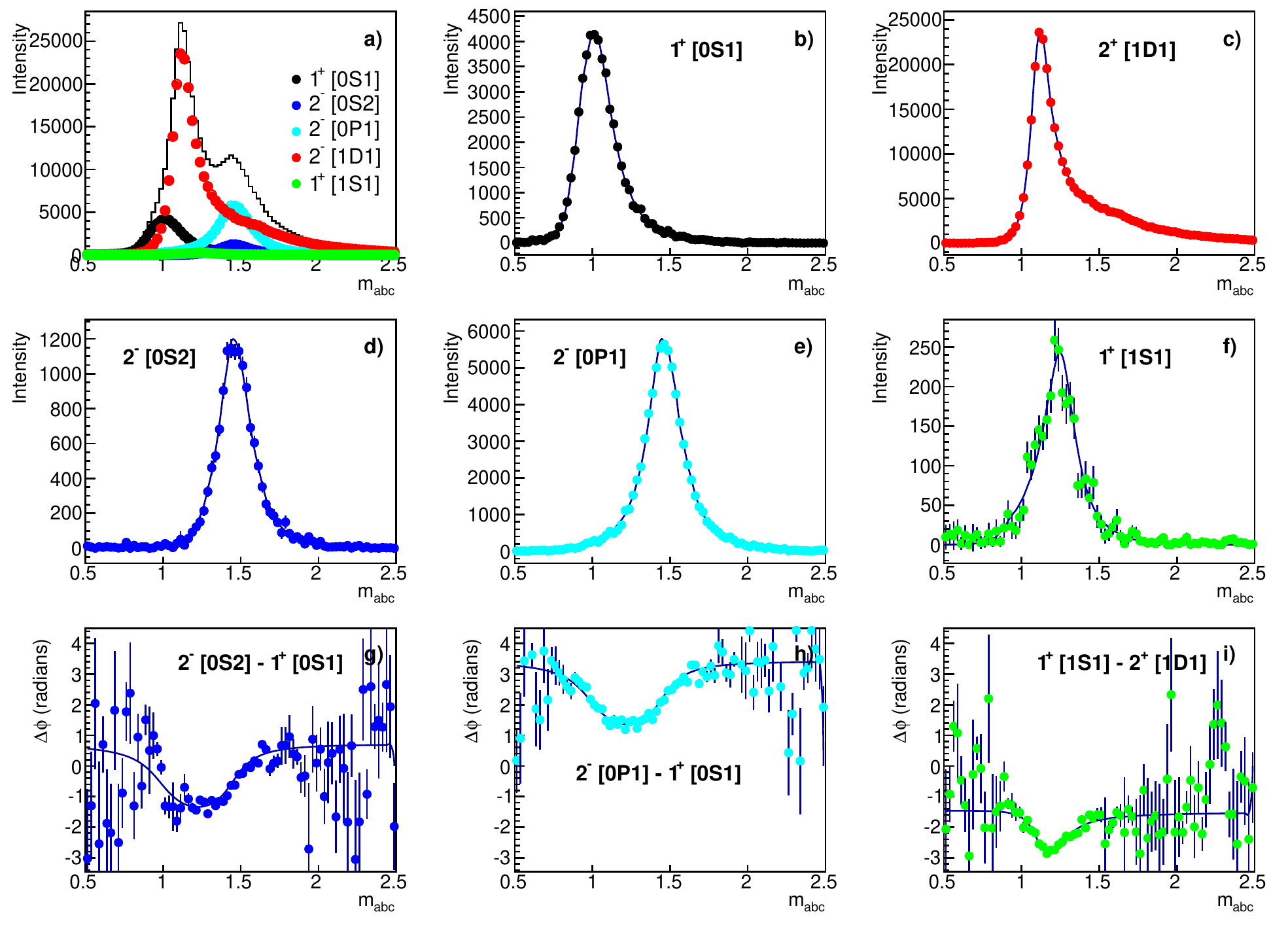}
                \caption{ Results of extended maximum likelihood fit with the true p.d.f. from Table~1: a)~-~f) intensities for each resonant amplitude in the true p.d.f. labeled $J^{P} [M l j]$ and g)~-~i) phase differences between each resonant wave and the corresponding reference wave with the same $M$.  Closed symbols represent the measured intensities and phase differences, while the true model values are indicated by the solid curves. }
                \label{fig:fitMass6wave}
        \end{center}
\end{figure}

In all cases the fit minimizes the quantity $-2\log{\mathcal{L}}$, where $\mathcal{L}$ is the extended likelihood function obtained from $|\mathcal{M}|^2$.  
Figure~\ref{fig:fitMass6wave} shows the results of a mass-independent fit where the p.d.f.\ used in the fit contains only the  amplitudes from the true model (\textit{i.e.}, those in Table~\ref{Table:ModelParams}).  The intensities of the individual amplitudes are accurately and precisely determined, and the clear resonance structure from the Breit-Wigner propagators is apparent in the extracted phase differences between resonant amplitudes in the coherent sum.  

%%%% Figure %%%%

\begin{figure}
        \begin{center}
                \includegraphics[width=1.0\textwidth]{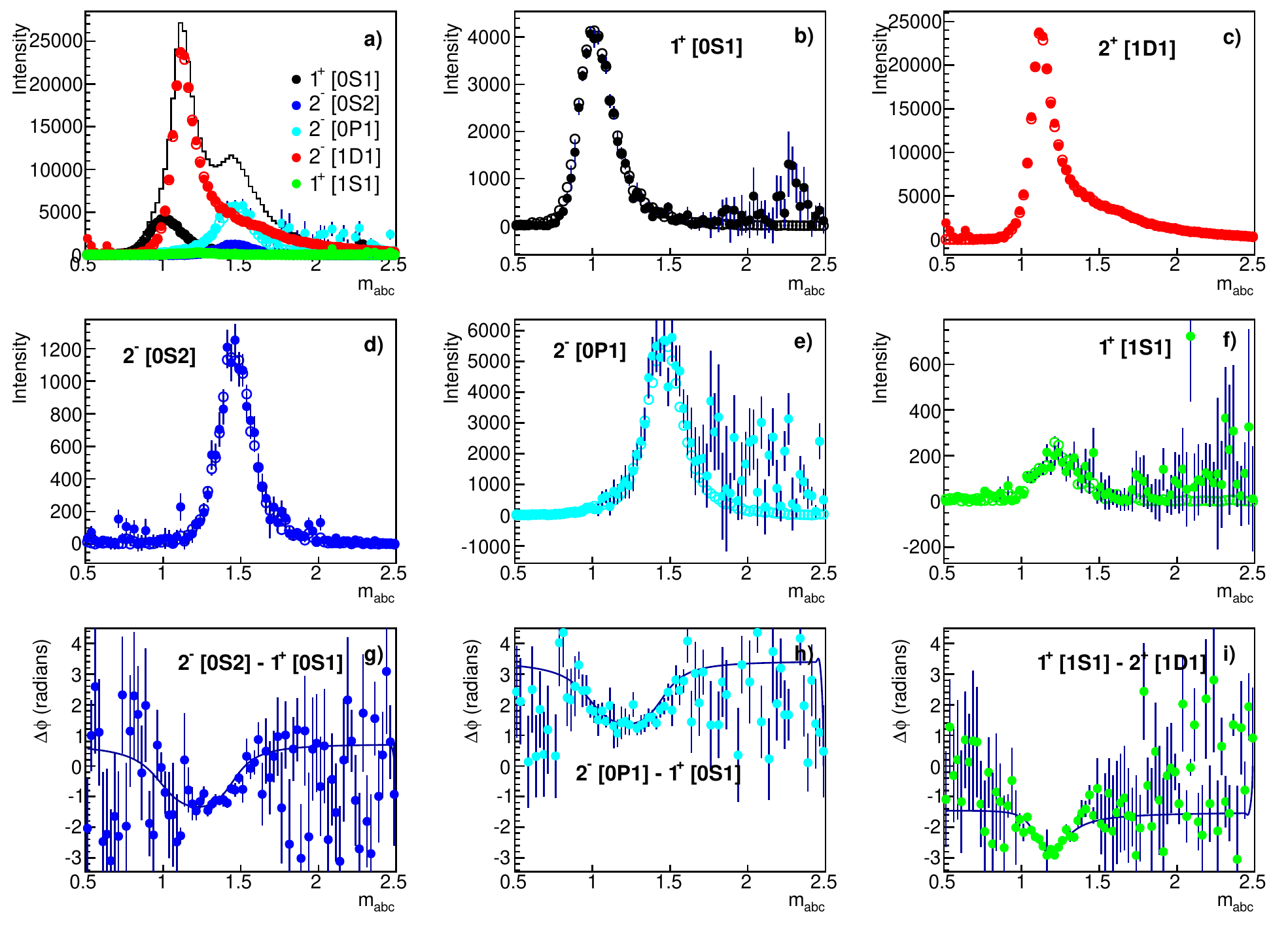}
                \caption{ Results of extended maximum likelihood fit with the full p.d.f., including the additional 35 extraneous amplitudes, and no LASSO regularization (\textit{i.e.} $\lambda=0$): a)~-~f) intensities for each resonant amplitude in the true p.d.f. labeled $J^{P} [M l j]$ and g)~-~i) phase differences between each resonant wave and the corresponding reference wave with the same $M$.  Closed symbols represent the measured intensities and phase differences for the full p.d.f., open symbols represent the intensities determined from the true p.d.f.\ fit in Fig.~2, and the curves indicate the true model values for the phase difference. }
                \label{fig:fitMass41wave}
        \end{center}
\end{figure}

In general, however, the true set of amplitudes to use is not known \textit{a priori}, and the full p.d.f.\ used in the fit could contain an arbitrarily large number of amplitudes.  
In this example, the p.d.f.\ is expanded to contain an additional 35 resonant amplitudes with additional allowed combinations of $J, M, j$ and $l$ listed in Table~\ref{Table:ModelParams_FullPDF}.
Figure~\ref{fig:fitMass41wave} shows the results of an extended maximum likelihood fit to the same data set but using the expanded set of amplitudes.
The performance of the fits has clearly deteriorated.  The variance of the estimators of the true model parameters has in many places greatly increased.  

\begin{figure}
        \begin{center}
                \includegraphics[width=0.5\textwidth]{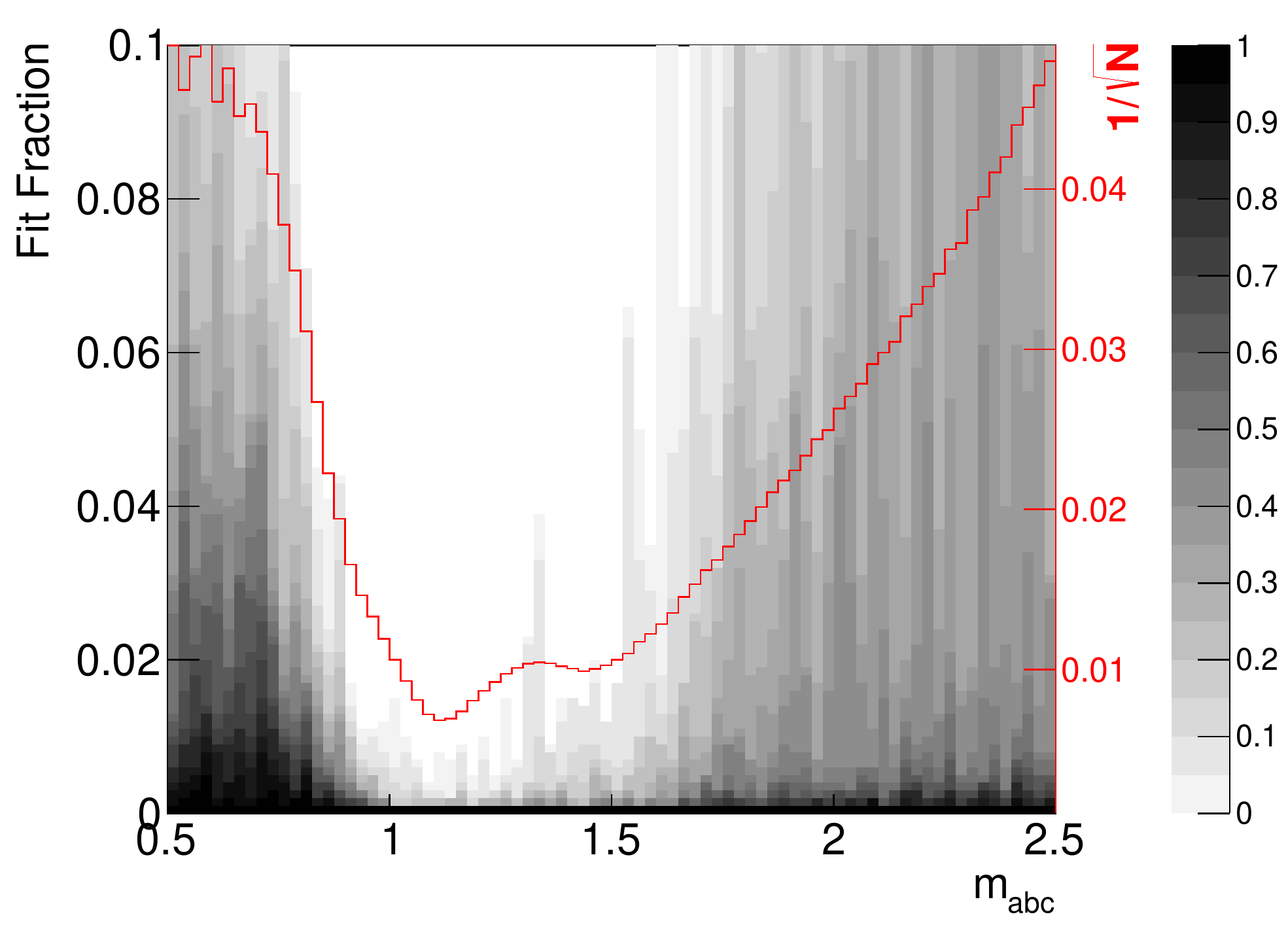}
                \caption{ Cumulative fraction of extraneous amplitudes with fit fraction greater than $y$-axis value {\em vs} $m_{abc}$. 
The (red) line shows $1/\sqrt{N}$, where $N$ is the total number of events in the $m_{abc}$ bin.  This is shown just to give a sense of statistical precision in the bin.
}
                \label{fig:cumFFmass}
        \end{center}
\end{figure}

Additionally, the fits incorrectly determine that some of the 35 extraneous amplitudes (those not in the true p.d.f.) make significant contributions.  The contribution from each resonant amplitude $k$ in the fit can be represented by the fit fraction, defined as $\int \left| a_k e^{i\phi_k} \mathcal{A}^M_k(\vec{x})\right|^2 d\vec{x}/\int |\mathcal{M}(\vec{x})|^2d\vec{x}$.  The cumulative fraction of extraneous amplitudes is then defined as the fraction of the 35 extraneous amplitudes with a fit fraction larger than a given threshold, which is shown in Fig.~\ref{fig:cumFFmass} {\em vs} $m_{abc}$.   In bins of $m_{abc}$ with low statistics, there are clearly sizable contributions from multiple extraneous amplitudes.

%%%% Table %%%%
\begin{table}[h!]
  \centering
  %\begin{small}
  \begin{tabular}{|c|c|c|c|c|c|c|}
    \hline
    $J^P$ & $M$ & $j$ & $l$ & Isobar mass & Isobar width & Isobar daughters \\
    \hline
    $0^-$ & 0 & 1 & 1 & 0.60 & 0.10 & $ab,ac$ \\
    $1^+$ & 0 & 1 & 0 & 0.60 & 0.10 & $ab,ac$ \\
    $1^+$ & 1 & 1 & 0 & 0.60 & 0.10 & $ab,ac$ \\
    $1^+$ & 0 & 1 & 2 & 0.60 & 0.10 & $ab,ac$ \\
    $1^+$ & 1 & 1 & 2 & 0.60 & 0.10 & $ab,ac$ \\
    $2^-$ & 1 & 1 & 1 & 0.60 & 0.10 & $ab,ac$ \\
    $2^+$ & 0 & 1 & 2 & 0.60 & 0.10 & $ab,ac$ \\
    $2^-$ & 0 & 1 & 3 & 0.60 & 0.10 & $ab,ac$ \\
    $2^-$ & 1 & 1 & 3 & 0.60 & 0.10 & $ab,ac$ \\
    $1^-$ & 0 & 1 & 1 & 0.60 & 0.10 & $ab,ac$ \\
    $1^-$ & 1 & 1 & 1 & 0.60 & 0.10 & $ab,ac$ \\
    $3^+$ & 0 & 1 & 2 & 0.60 & 0.10 & $ab,ac$ \\
    $3^+$ & 1 & 1 & 2 & 0.60 & 0.10 & $ab,ac$ \\
    $0^-$ & 0 & 1 & 1 & 0.75 & 0.10 & $bc$ \\
    $1^+$ & 0 & 1 & 2 & 0.75 & 0.10 & $bc$ \\
    $1^+$ & 1 & 1 & 2 & 0.75 & 0.10 & $bc$ \\
    $2^-$ & 0 & 1 & 1 & 0.75 & 0.10 & $bc$ \\
    $2^-$ & 1 & 1 & 1 & 0.75 & 0.10 & $bc$ \\
    $2^+$ & 0 & 1 & 2 & 0.75 & 0.10 & $bc$ \\
    $2^+$ & 1 & 1 & 2 & 0.75 & 0.10 & $bc$ \\
    $2^-$ & 0 & 1 & 3 & 0.75 & 0.10 & $bc$ \\
    $2^-$ & 1 & 1 & 3 & 0.75 & 0.10 & $bc$ \\
    $1^-$ & 0 & 1 & 1 & 0.75 & 0.10 & $bc$ \\
    $1^-$ & 1 & 1 & 1 & 0.75 & 0.10 & $bc$ \\
    $3^+$ & 0 & 1 & 2 & 0.75 & 0.10 & $bc$ \\
    $3^+$ & 1 & 2 & 2 & 0.75 & 0.10 & $bc$ \\
    $1^+$ & 0 & 2 & 1 & 1.10 & 0.15 & $bc$ \\
    $1^+$ & 1 & 2 & 1 & 1.10 & 0.15 & $bc$ \\
    $2^-$ & 1 & 2 & 0 & 1.10 & 0.15 & $bc$ \\
    $2^-$ & 0 & 2 & 2 & 1.10 & 0.15 & $bc$ \\
    $2^-$ & 1 & 2 & 2 & 1.10 & 0.15 & $bc$ \\
    $2^+$ & 0 & 2 & 1 & 1.10 & 0.15 & $bc$ \\
    $2^+$ & 1 & 2 & 1 & 1.10 & 0.15 & $bc$ \\
    $3^+$ & 0 & 2 & 1 & 1.10 & 0.15 & $bc$ \\
    $3^+$ & 1 & 2 & 1 & 1.10 & 0.15 & $bc$ \\
    \hline
  \end{tabular}
  %\end{small}
  \caption{ The set of extraneous resonant amplitudes and properties used in the full p.d.f.\ for the Dalitz-plot example.  The symbols are: $J^P$ and $M$ the spin-parity and spin projection of $X$;  $j$ is the isobar spin; and $l$ is the orbital angular momentum between the isobar and the bachelor.}
  \label{Table:ModelParams_FullPDF}
\end{table}

\clearpage

%%%% Figure %%%%

%%%% Figure %%%%

%%%%%%%
%  Method  %
%%%%%%%
\section{Method}
\label{Sec:Method}

Identifying a "good" model consists not only of adequately describing the data, {\em e.g.}, using one of the multivariate goodness-of-fit methods described in detail in Ref.~\cite{GOF},
but also in properly selecting the subset of amplitudes to include.
A common approach taken to address this problem involves stepwise addition or subtraction of amplitudes from the model (for a detailed description of such methods, see Ref.~\cite{Miller_Subsets}); however, this approach is known to often produce an over-simplified model\cite{Harrell}. Furthermore, the stepwise approach requires vast CPU resources for large data sets.

An alternative approach is to regulate the regression procedure.
A popular choice (outside of particle physics) is the Least Absolute Shrinkage and Selection Operator (LASSO)\cite{LASSO_Tib_96}.
The LASSO imposes a so-called $L_{1}$-type regularization condition on the sum of the absolute values of the regression parameters, which tends to produce sparse models; {\em i.e.}, the LASSO tends to produce models with few significantly non-zero regression parameters.
The use of the sum of the squared values of the regression parameters is referred to as $L_2$-type regularization and is used, {\em e.g.}, in so-called Ridge regression. 
Since Ridge regression uses the squared parameter values, applying it to an amplitude anlaysis would remove excessive interference; however, it would not drive {\em unphysical} model parameters to zero, and so is not as well suited to the task studied in this paper.

Applying the LASSO to this analysis involves modifying the function that is minimized in the fit as follows:
\begin{equation}
\label{Eqn:LASSOLikelihood}
-2\log{\mathcal{L}} + \lambda \left[ \sum_{i,M} \sqrt{ \int \left| a_i e^{i\phi_i} \mathcal{A}^M_i(\vec{x})\right|^2 d\vec{x} } + \sqrt{\int |a_{nr}|^2 d\vec{x}} \right],
\end{equation}
where $\lambda$ is the LASSO regularization parameter.
Our choice of LASSO penalty term does not depend on the relative normalization of the $\mathcal{A}(\vec{x})$ terms and has the advantage that it requires almost no modification to existing fitting software\footnote{Note that the penalty term is calculated using only information already required to calculate the nominal function minimized in the fit.}.

A proper choice of $\lambda$ will reduce the number of amplitudes contributing to the fit result, similar to what is done explicitly with {\em ad hoc} procedures when not using regularization.  However, the subset of amplitudes selected by the LASSO is that which minimizes Eq.~\ref{Eqn:LASSOLikelihood} which does not require independently running fits with all combinations of possible sets of amplitudes.  
Roughly speaking, the LASSO can be said to find the optimal model to describe the data with the parameter $\lambda$ setting the desired maximum amount of model complexity, but not explicitly setting a fixed number of coefficients to zero.   
Note that when $\lambda \rightarrow 0$ the LASSO solution reduces to the nominal un-penalized one. 
Conversely, when $\lambda \rightarrow \infty$ the LASSO produces a model where few (if any) coefficients are nonzero. 
How to choose the value for $\lambda$ is discussed below.

\begin{figure}
        \begin{center}
                \includegraphics[width=1.0\textwidth]{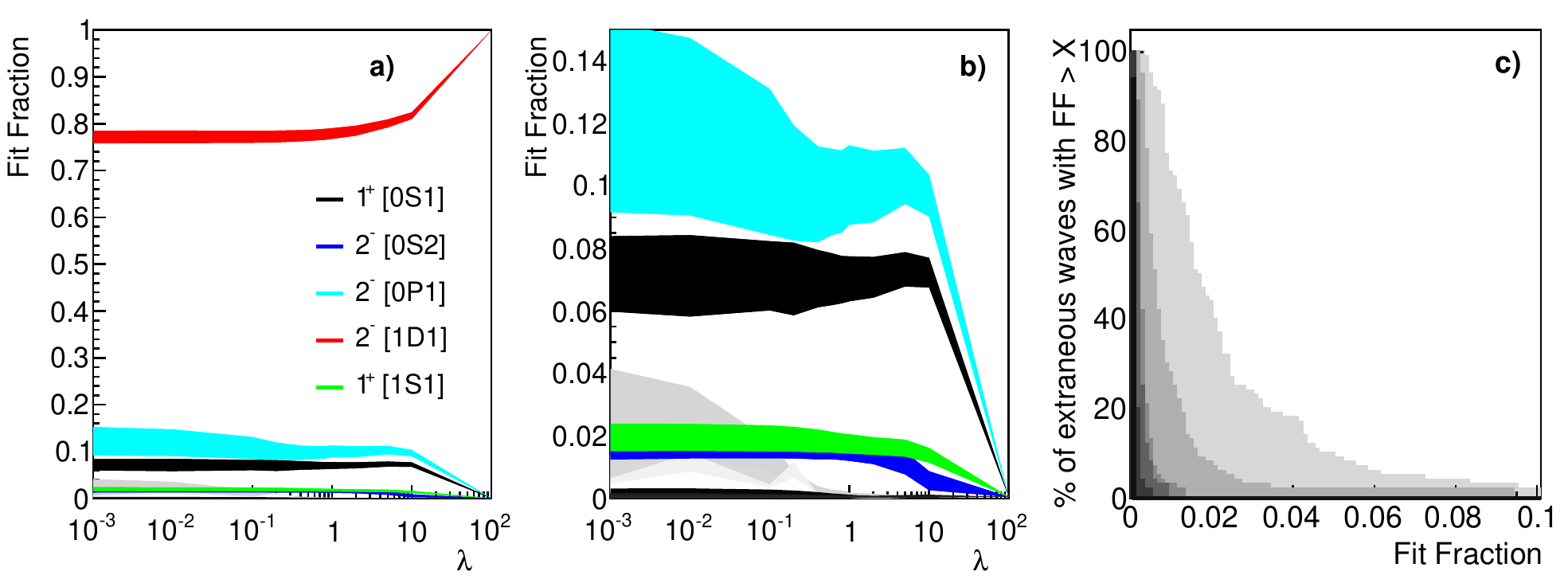}
                \caption{ Results of fitting an ensemble of 100 data sets in a single mass bin with $m_{abc} = 1.25$.  a) Fit fractions for each of the resonant amplitudes in the full p.d.f.\ as a function of $\lambda$, where the width of each band represents the standard deviation obtained from the 100 data samples in the ensemble.  b) is the same as a), but with a zoomed $y$-axis to better display the growing contributions of the extraneous resonant amplitudes as $\lambda\rightarrow 0$.  c) Percentage of extraneous resonant amplitudes with fit fraction > $X$ for $\lambda$ = 0, 0.1, 0.4, 0.8, 2, 10 (increasing from light grey to black); this demonstrates the reduction in extraneous amplitudes contributing to the fit as $\lambda$ is increased. }
                \label{fig:cumFFensemble}
        \end{center}
\end{figure}

Figure~\ref{fig:cumFFensemble} shows the results of fitting an ensemble of 100 data sets in a single $m_{abc}$ bin ($m_{abc} = 1.25$) using values of $\lambda$ from 0 to 100.  
As $\lambda$ is increased from zero the contributions due to extraneous amplitudes decrease and the precision on the parameters of the amplitudes in the true model increases.  As $\lambda$ approaches 100, however, even the fit fractions of the amplitudes that are in the true model are driven towards zero. 
Knowing the true model, it is clear to see that the optimal value of $\lambda$ is $\mathcal{O}(1)$ in this example.

%%%%%

The final puzzle piece then is to determine how much complexity in the model is optimal.
This involves balancing improvements in the nominal minimization quantity ($-2\log{\mathcal{L}}$) with additional complexity in the model.  There are many methods in the literature for choosing $\lambda$ using the data sample itself.  
Two considered in this paper are the Akaike and Bayesian information criteria (AIC~\cite{AIC_74} and BIC~\cite{BIC_78}):
\begin{equation}
\textrm{AIC}(\lambda) = -2\log{\mathcal{L}} + 2r, \qquad \textrm{BIC}(\lambda) = -2\log{\mathcal{L}} + r\log{n},
\label{eqn:aicbic}
\end{equation}  
where $n$ is the number of events in the data set and $r$ is related to the number of amplitudes in the model that have a fit fraction sizably different from zero ($r$ is discussed in detail in the examples below).
Therefore, both AIC and BIC take into account both improvements in the goodness-of-fit of the data and complexity of the model.  In our studies below both AIC and BIC perform well.

Combining the LASSO method with a chosen information criterion permits performing the amplitude selection procedure quickly using the data sample itself.  
The full procedure to be applied to a given data set is as follows:
\begin{itemize}
\item choose the set of all possible amplitudes to be considered for inclusion in the model;
\item for any $\lambda$, minimize Eq.~\ref{Eqn:LASSOLikelihood} to obtain $-2\log{\mathcal{L}}$ and the regression parameters;
\item calculate AIC($\lambda$) or BIC($\lambda$) with determination of $r$ discussed below;
\item scan $\lambda$ values to find the one that minimizes AIC or BIC.
\end{itemize}
The value of $\lambda$ that minimizes the chosen information criterion, $\hat{\lambda}$, is selected.  
Since penalty terms are added to the minimization quantity, a scan of the profile likelihood may no longer produce proper confidence intervals for the regression parameters.    
In this paper the variances in regression parameters are determined using the bootstrap~\cite{bootstrap}.  Bootstrap data sets are resampled from the original, and the entire procedure -- including optimizing $\lambda$ -- is repeated on each copy data set independently.    
An alternative approach is to use the LASSO to choose the model, and then rerun the fit using the selected waves but no penalty term.

%%%% Figure %%%%

In this study, we have chosen $r$ to be the number of resonant amplitudes with a fit fraction larger than $10^{-3}$.  Since each resonant amplitude has both an unknown magnitude and a phase, or equivalently real and imaginary factors, we also considered taking $r$ to be twice the number of resonant amplitudes with a fit fraction greater than $10^{-3}$ but we see almost no change in the choice of optimal $\lambda$.  Furthermore, we tried increasing and decreasing the $10^{-3}$ threshold by a factor of 10 but also see very little change in the choice of optimal $\lambda$.  

\begin{figure}
        \begin{center}
                \includegraphics[width=1.0\textwidth]{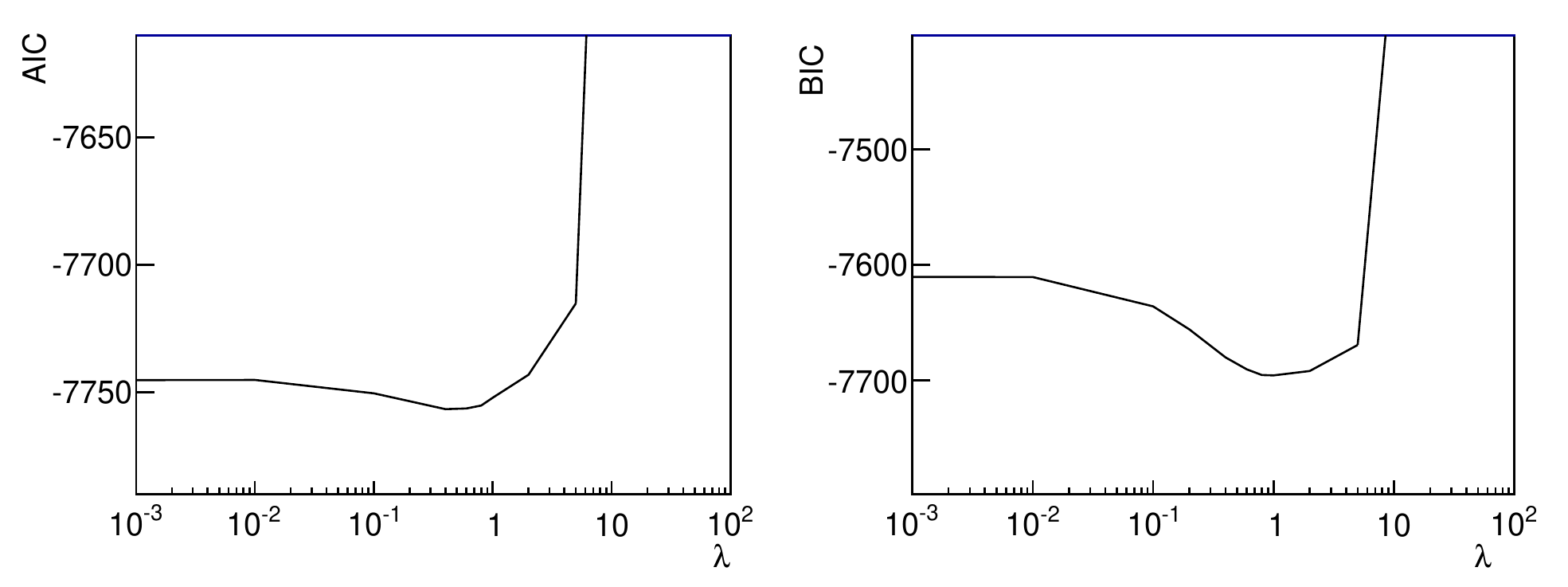}
	       \caption{ Average a) AIC and b) BIC as a function of $\lambda$ for the ensemble of 100 independent samples with $m_{abc} = 1.25$.}
                \label{fig:BIC}
        \end{center}
\end{figure}

\begin{figure}
        \begin{center}
                \includegraphics[width=1.0\textwidth]{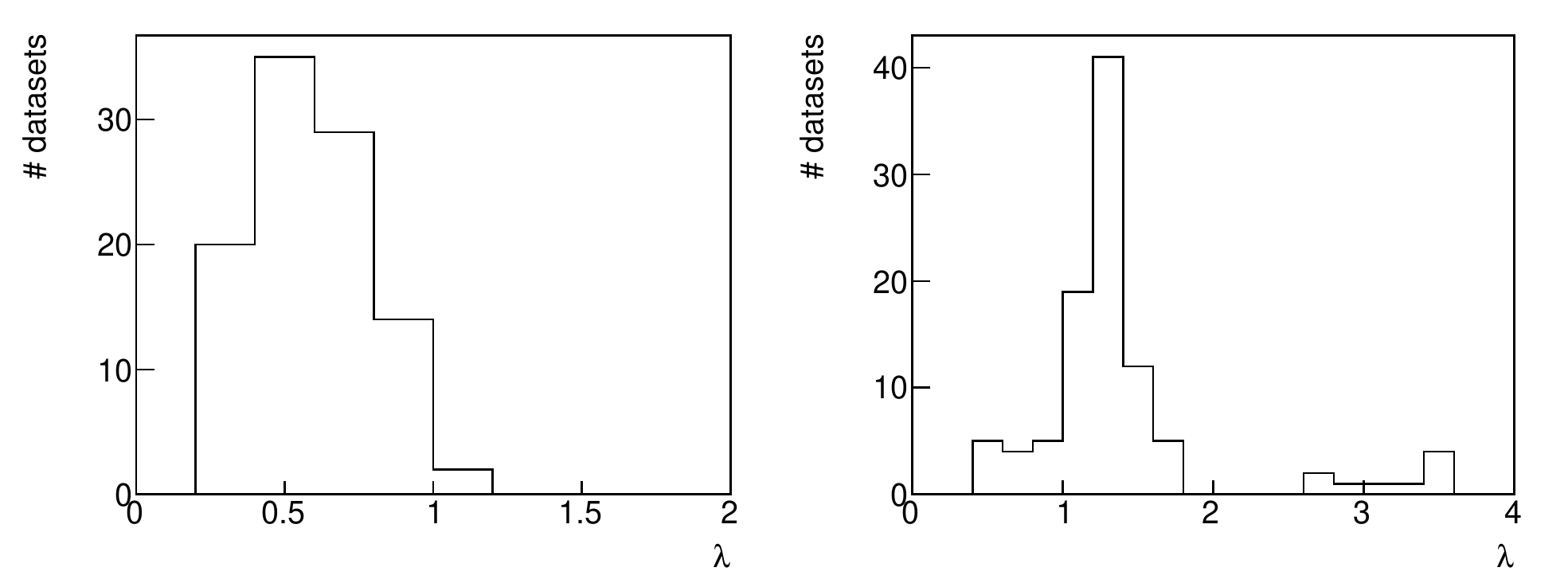}
	       \caption{ Distribution of $\lambda$ values which minimize a) AIC and b) BIC for the ensemble of 100 independent samples with $m_{abc} = 1.25$.}
                \label{fig:BIClambda}
        \end{center}
\end{figure}

Figure~\ref{fig:BIC} shows the average AIC and BIC from the ensemble of datasets used to produce Fig.~\ref{fig:cumFFensemble} as a function of $\lambda$.  The optimal AIC and BIC both occur near $\lambda=1$.  By construction, the BIC method chooses a slightly larger $\lambda$ value and, thus, on average a less complex model.  
Figure~\ref{fig:BIClambda} shows the optimal value of $\lambda$ chosen for 100 independent data samples. 
In a real analysis, we suggest to try both AIC and BIC and use the difference in results obtained as part of the estimate for the systematic uncertainty due to model selection.

%%%% Figure %%%%

%%%% Figure %%%%

\begin{figure}
        \begin{center}
                \includegraphics[width=1.0\textwidth]{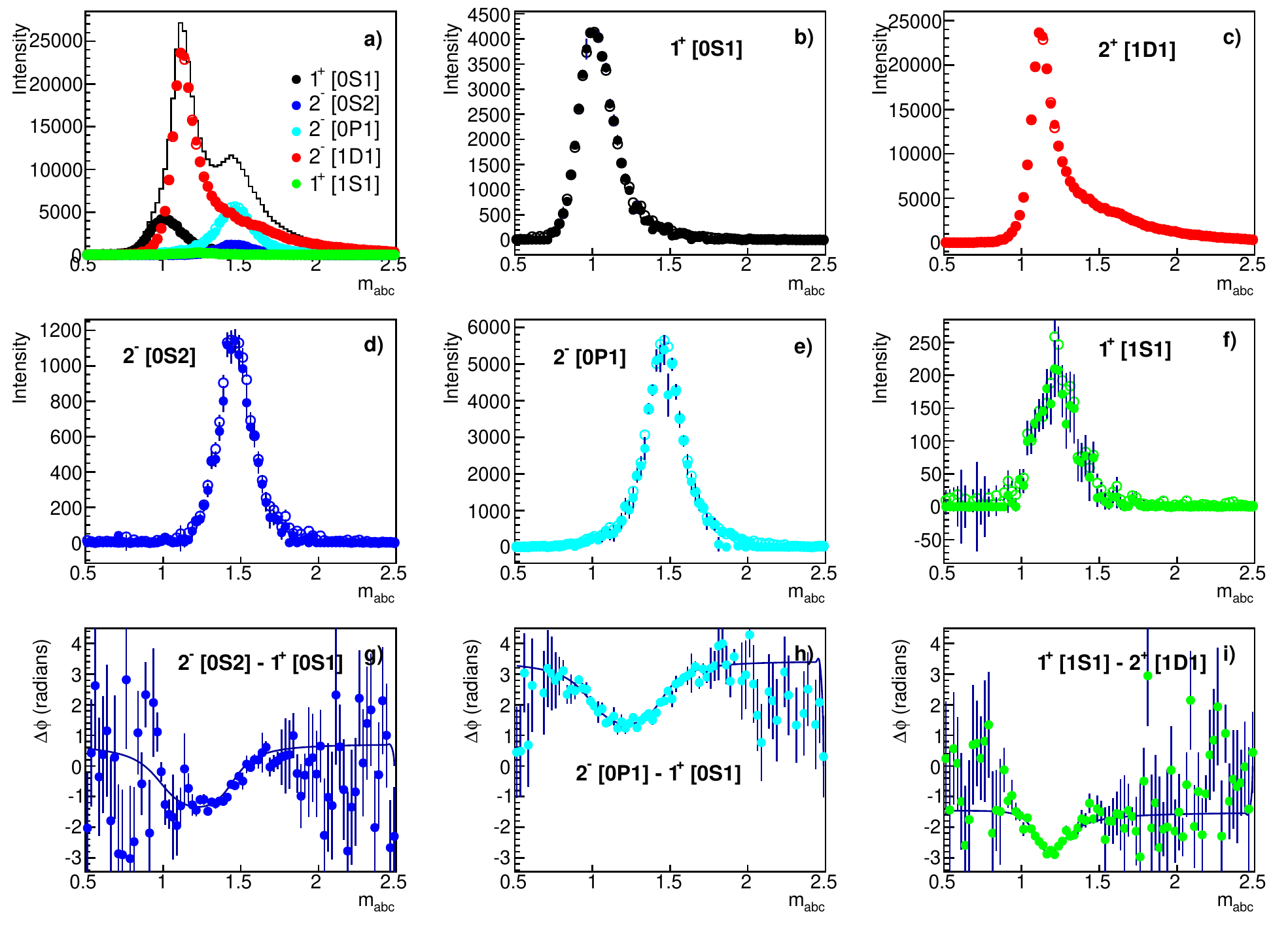}
                \caption{ Results of extended maximum likelihood fit with the full p.d.f., including the additional 35 extraneous amplitudes, and using BIC to select the optimal $\lambda$ for each $m_{abc}$ bin: a)~-~f) intensities for each resonant amplitudes in the true p.d.f. labeled $J^{P} [M l j]$ and g)~-~i) phase differences between each resonant wave and the corresponding reference wave with the same $M$.  Closed symbols represent the measured intensities and phase differences for the full p.d.f.\ using the LASSO, open symbols represent the intensities determined from the true p.d.f.\ fit in Fig.~2, and the curves indicate the true model values for the phase difference.  The agreement between the fit results using the true model p.d.f.\ {\em vs} the full model using BIC-LASSO is good enough that is difficult to see both results in these plots since they overlap.}
                \label{fig:fitMassBIC}
        \end{center}
\end{figure}

\begin{figure}[]
        \begin{center}
                \includegraphics[width=0.5\textwidth]{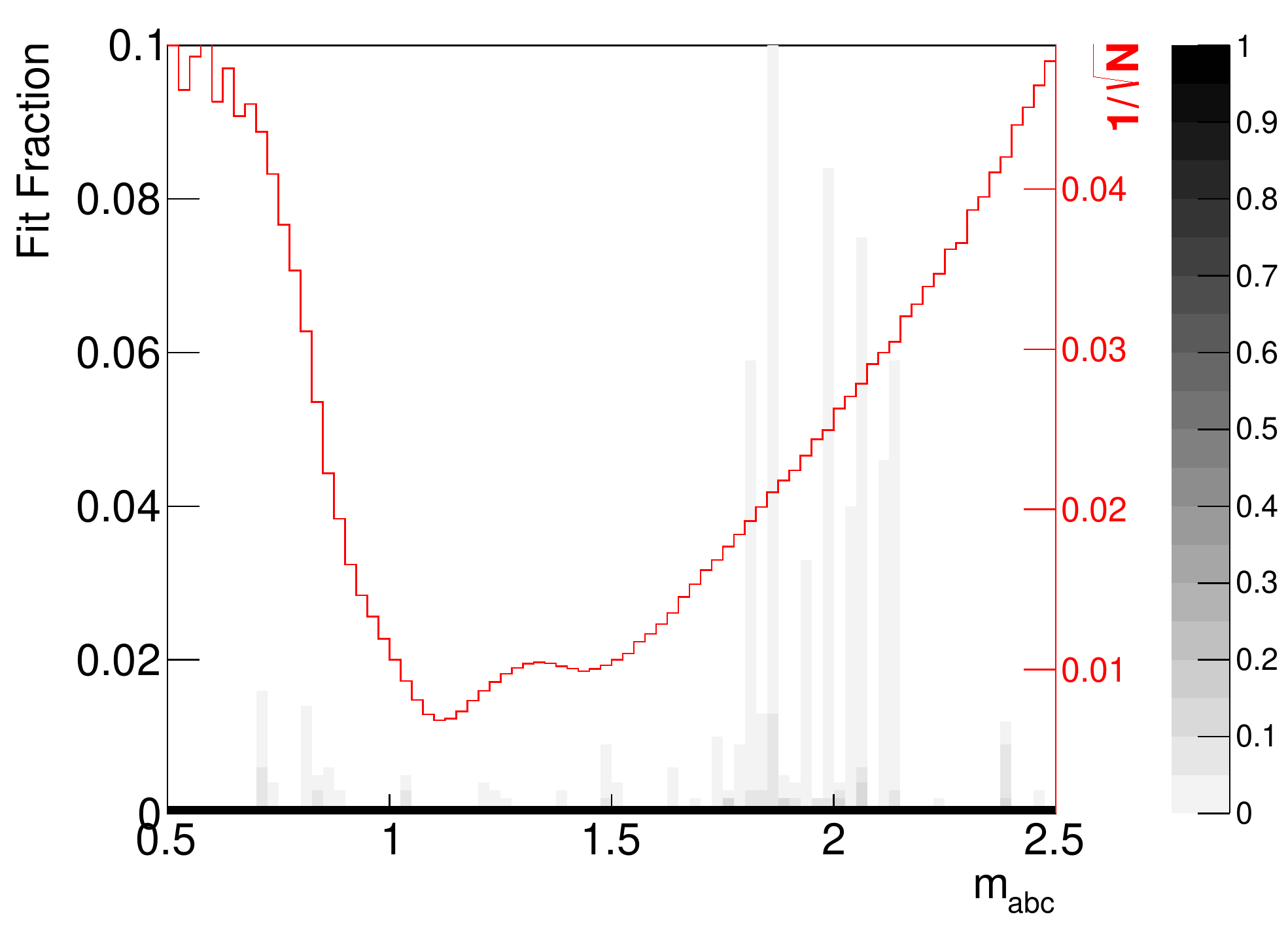}
                \caption{ Cumulative fraction of extraneous amplitudes with fit fraction greater than $y$-axis value {\em vs} $m_{abc}$ using BIC to select the optimal $\lambda$ for each $m_{abc}$ bin. 
The (red) line shows $1/\sqrt{N}$, where $N$ is the total number of events in the $m_{abc}$ bin.  This is shown just to give a sense of statistical precision in the bin.
}
                \label{fig:cumFFmassBIC}
        \end{center}
\end{figure}

Figure~\ref{fig:fitMassBIC} shows the results of the mass-independent fit where the BIC-LASSO procedure is used to select the optimal $\lambda$ in each $m_{abc}$ bin separately. 
The amplitudes from the true p.d.f.\ are more accurately determined in this case, and as seen in Fig.~\ref{fig:cumFFmassBIC} the number of extraneous amplitudes with non-zero fit fractions is dramatically reduced.
For this Dalitz-plot analysis, using the BIC-LASSO produces results similar to those of using only the true model p.d.f.\ which, of course, is not known in the analysis of real data. 
This is an impressive result given that we allowed the initial set of amplitudes considered to be large.  An important point to make here is that this model simplification was achieved without direct human intervention during the process.

%%%% Figure %%%%

%%%% Figure %%%%

To evaluate the accuracy of the measured intensities from the fit, we study the  measured fit fractions in the $m_{abc}$ bin with the largest intensity in the true p.d.f.\ for an ensemble of 1000 datasets
% in Figs.~\ref{fig:FitFractions} and~\ref{fig:FitFractionsBIC} 
without using the LASSO procedure (i.e. $\lambda = 0$) and with the LASSO procedure using BIC to select the best $\lambda$ for each data sample and $m_{abc}$ mass bin, respectively.  Table~\ref{Table:FitFractionsModel1} shows that the  measured fit fractions are more precisely determined and less biased when using the BIC-LASSO procedure.

%%%% Table %%%%
\begin{table}[]
  \centering
  \begin{tabular}{|c|c|c|c|c|}
    \hline
    $J^P [Mlj]$ & Generated & True p.d.f. & No LASSO & BIC-LASSO  \\
    \hline
    $1^+ [0S1]$ & 41.74  & 41.78 $\pm$ 0.76 & 40.37 $\pm$ 1.73 & 41.83 $\pm$ 0.86  \\
    $2^- [0S2]$ & 10.55 & 10.12 $\pm$ 0.52 & 10.58 $\pm$ 0.96 & 10.11 $\pm$ 0.71 \\
    $2^- [0P1]$ & 49.21 & 48.67 $\pm$ 0.91 & 53.40 $\pm$ 5.87 & 48.04 $\pm$ 2.62 \\
    $2^- [1D1]$ & 86.58 & 86.64 $\pm$ 0.40 & 86.47 $\pm$ 1.01 & 86.5 $\pm$ 0.77 \\
    $1^+ [1S1]$ & 1.77 & 1.90 $\pm$ 0.23 & 1.62 $\pm$ 0.42 & 1.60 $\pm$ 0.40 \\
    \hline
  \end{tabular}
  \caption{ Mean and standard deviation of the measured fit fractions for Model I. }
  \label{Table:FitFractionsModel1}
\end{table}

%%%%%%%%%%%%%%%
%%      Alternate model      %%
%%%%%%%%%%%%%%%

In reality the true model is likely to contain many non-zero but small resonant contributions that the analyst is not interested in.  To study this case,  an alternate model with six additional resonant amplitudes with fit fractions all below 1\% (see Table~\ref{Table:AltModelParams}) is used to study the effectiveness of the BIC-LASSO procedure when there are a number of low intensity amplitudes in the true p.d.f.  Throughout the remainder of the paper this alternate model is referred to as Model II.  The same measured fit fraction distributions as in Table~\ref{Table:FitFractionsModel1}
are shown here in 
Table~\ref{Table:FitFractionsModel2}
for Model II.  The amplitudes in the true p.d.f.\ remain unbiased when using BIC-LASSO even in the presence of many additional low intensity amplitudes.

%%%% Table %%%%
\begin{table}[]
  \centering
  \begin{tabular}{|c|c|c|c|c|c|c|c|c|}
    \hline
    $J^P$ & $m_X$ & $\Gamma_X$ & $M$ & $j$ & $l$ & Isobar mass & Isobar width & Isobar daughters \\
    \hline
    $1^-$ & 1.55 & 0.40 & 0 & 1 & 1 & 0.75 & 0.10 & $bc$ \\
    $1^-$ & 1.55 & 0.40 & 0 & 1 & 1 & 0.60 & 0.10 & $ab, ac$ \\
    $3^+$ & 1.65 & 0.50 & 0 & 1 & 2 & 0.75 & 0.10 & $bc$ \\
    $2^-$ & 1.45 & 0.25 & 1 & 1 & 1 & 0.60 & 0.10 & $ab, ac$ \\
    $3^+$ & 1.65 & 0.50 & 1 & 1 & 2 & 0.75 & 0.10 & $bc$ \\
    $3^+$ & 1.65 & 0.50 & 1 & 2 & 1 & 1.10 & 0.15 & $bc$ \\
    \hline
  \end{tabular}
  \caption{ The set of resonant terms and properties (in addition to those in Table 1) used in the true p.d.f.\ for Model II.  The symbols are: $J^P$, $m_X$, $\Gamma_X$, $M$ the spin-parity, mass, width and spin projection of $X$;  $j$ is the isobar spin; and $l$ is the orbital angular momentum between the isobar and the bachelor.}
  \label{Table:AltModelParams}
\end{table}

%%%% Table %%%%
\begin{table}[]
  \centering
  \begin{tabular}{|c|c|c|c|c|}
    \hline
    $J^P [Mlj]$ & Generated & True p.d.f. & No LASSO & BIC-LASSO  \\
    \hline
    $1^+ [0S1]$ & 41.76 & 41.11 $\pm$ 0.76 & 39.79 $\pm$ 1.19 & 40.75 $\pm$ 0.79 \\
    $2^- [0S2]$ & 9.99 & 9.99 $\pm$ 0.47 & 10.89 $\pm$ 1.01 & 10.65 $\pm$ 0.83 \\
    $2^- [0P1]$ & 48.80 & 48.50 $\pm$ 1.00 & 53.99 $\pm$ 7.41 & 47.37 $\pm$ 4.54 \\
    $2^- [1D1]$ & 86.46 & 86.41 $\pm$ 0.40 & 87.22 $\pm$ 0.78 & 87.13 $\pm$ 0.76 \\
    $1^+ [1S1]$ & 1.98 & 1.95 $\pm$ 0.23 &1.80 $\pm$ 0.37 & 1.75 $\pm$ 0.36 \\
    \hline
  \end{tabular}
  \caption{ Mean and standard deviation of the measured fit fractions for Model II. }
  \label{Table:FitFractionsModel2}
\end{table}

%%%%%%%%%%%%%%%
%% 	     Resonance ID	 %%
%%%%%%%%%%%%%%%
\section{Identification of resonances}

\begin{figure}
        \begin{center}
                \includegraphics[width=1.0\textwidth]{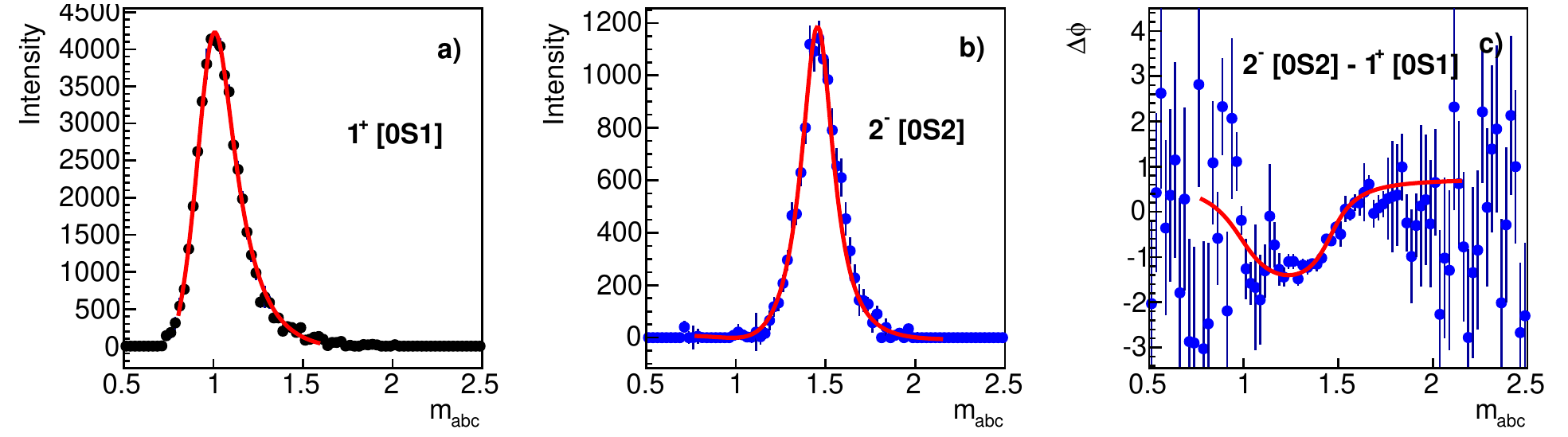}
                \caption{ Resonant amplitude intensity for a) $1^+ [0S1]$ and b) $2^- [0S2]$ and c) phase difference between the two amplitudes.  The red curves are a simultaneous fit to the three distributions described in the text. }
                \label{fig:resonanceID}
        \end{center}
\end{figure}

In this section we present an outline for how to develop a procedure for identifying resonant structures in the results of the mass-independent fit from the previous section.  The goal of such a procedure is to search for new $X$ resonances or measure the properties of known resonances in the $m_{abc}$ mass spectrum.  A resonance is typically observed through both a significant intensity and corresponding phase motion in a particular amplitude.  For example, Fig.~\ref{fig:resonanceID} shows a significant intensity for the $2^- [0S2]$ state and resonant-like phase motion relative to the $1^+ [0S1]$ amplitude (these results are obtained from the BIC-LASSO regulated mass-independent fits described in the previous section).  
The purpose of this section is to show that if such a procedure can be developed prior to examining the data --- recall that employing the BIC-LASSO permits defining the model in each $m_{abc}$ bin using the data itself --- it would be possible to define the statistical significance of each resonance.  

There are a number of difficulties that must be addressed by any procedure meant to identify resonance contributions:
\begin{enumerate}
\item The functional form of the $m_{abc}$ dependence of the intensity and phase for a resonant amplitude is likely not known {\em a priori}.  Resonances are often well described by the terms included in our model amplitudes; however, for certain cases, {\em e.g.}\ when multiple resonances with the same quantum numbers overlap in $m_{abc}$, this prescription is not valid.  
\item Even in the best case scenario, there will be unknown parameters in the functional form discussed in (1) that invalidate the use of Wilks' Theorem~\cite{wilks}.  Obtaining the significance will require generating a large ensemble of pseudo data sets.  
\item There may be multiple resonances with the same quantum numbers contributing to different regions of $m_{abc}$; therefore, one either needs to scan in $m_{abc}$ and consider restricted regions, or perform a fit which allows for an arbitrary number of resonance contributions.
\item Since an analysis will typically look for resonance contributions in many different amplitudes, each with an unknown mass, the statistical significance must account for the so-called trials factor~\cite{gross}.  
\end{enumerate}
Developing a general procedure to identify resonances will be difficult, but it is certainly worthy of investigation.

%%%% Figure %%%%

In this section, we provide an {\em ad hoc} procedure that works well for this specific toy analysis.  We stress that our goal here is to demonstrate that if such a procedure is known {\em a priori},  it can be used in combination with the BIC-LASSO procedure to determine the statistical significance of each resonance.  
To identify a resonance in, {\em e.g.}, the $2^- [0S2]$ amplitude, there are three relevant distributions to consider: 1) the intensity of the $1^+ [0S1]$ reference amplitude, 2) the intensity of the $2^- [0S2]$ amplitude of interest, and 3) the phase difference $\Delta\phi$ between the two amplitudes.  The functional form discussed in point (1) above is taken directly from the true model p.d.f., but with the resonance parameters treated as unknowns.  

%In this study the resonant intensities are modeled by the norm of a complex function
%\begin{equation}
%f(m_{abc}; \vec{c}) \times BW \left(m_{abc}; m_{X}, \Gamma_{X}\right),
%\label{eqn:resonanceID}
%\end{equation}
%\noindent where 
%$f$ is a real-valued function with (possibly unknown) parameters $\vec{c}$ that parametrizes the available phase space, vertex factors, {\em etc.}, 
%and $BW$ is the Breit-Wigner distribution with resonance mass $m_{X}$ and width $\Gamma_{X}$.  
%The measured phase difference is modeled by the phase difference between the Breit-Wigner functions for the reference amplitude and the amplitude of interest in Eq.~\ref{eqn:resonanceID}.  
%The functional form used here is an {\em ad hoc} choice that describes the generated data. 
%In a real analysis the form of Eq.~\ref{eqn:resonanceID} used in this procedure would need to be carefully studied.

The three distributions in Fig.~\ref{fig:resonanceID} are fit simultaneously using $\chi^2$ minimization\footnote{We use the notation of $\chi^2$ here since the minimization quantity is constructed as the sum of the square of the (observed-expected)/uncertainty in each bin.  This quantity, however, is not expected to follow a $\chi^2$ distribution, and Wilks' Theorem is not expected to be valid as discussed in point (2) above.}, where the reference amplitude is assumed to be a known state and, therefore, its mass and width are fixed in the fit.  
Point (3) is addressed by scanning in $m_{abc}$ for each amplitude, and restricting the fit range to a maximum of $\pm0.5$ and minimum of  6 $m_{abc}$ bins.  This approach works well for the resonances included in our model; in a real analysis, one would need to determine these limits by other means.
The actual range used for each resonance fit is derived by requiring consecutive bins with intensity more than 2$\sigma$ from zero.  
%If this results in a range less than the minimum, the fit is discarded.  
 Tables~\ref{Table:Masses} and~\ref{Table:Widths} report the measured masses and widths of the Breit-Wigner fits for the non-reference amplitudes in the true p.d.f.\  The LASSO greatly improves the resolution on the resonance parameters.

%%%% Table %%%%
\begin{table}[]
  \centering
  \begin{tabular}{|c|c|c|c|c|c|}
    \hline
    & & \multicolumn{2}{c|}{Model I} & \multicolumn{2}{c|}{Model II} \\
    \hline
    $J^P [Mlj]$ & Generated & No LASSO & BIC-LASSO & No LASSO & BIC-LASSO \\
    \hline
    $2^- [0S2]$ & 1.45 & 1.46 $\pm$ 0.08 & 1.45 $\pm$ 0.01 & 1.47 $\pm$ 0.04 & 1.45 $\pm$ 0.01 \\
    $2^- [0P1]$ & 1.45 & 1.44 $\pm$ 0.07 & 1.43 $\pm$ 0.01 & 1.43 $\pm$ 0.09 & 1.45 $\pm$ 0.04 \\
    $1^+ [1S1]$ & 1.25 & 1.26 $\pm$ 0.03 & 1.25 $\pm$ 0.01 & 1.26 $\pm$ 0.04 & 1.25 $\pm$ 0.01 \\
    \hline
  \end{tabular}
  \caption{ Mean and standard deviation of the measured $m_X$ distributions extracted from the Breit-Wigner fits. }
  \label{Table:Masses}
\end{table}

%%%% Table %%%%
\begin{table}[]
  \centering
  \begin{tabular}{|c|c|c|c|c|c|}
  \hline
    & & \multicolumn{2}{c|}{Model I} & \multicolumn{2}{c|}{Model II} \\
    \hline
    $J^P [Mlj]$ & Generated & No LASSO & BIC-LASSO & No LASSO & BIC-LASSO \\
    \hline
    $2^- [0S2]$ & 0.25 & 0.25 $\pm$ 0.03 & 0.25 $\pm$ 0.01 & 0.25 $\pm$ 0.02 & 0.25 $\pm$ 0.01 \\
    $2^- [0P1]$ & 0.25& 0.27 $\pm$ 0.03 & 0.24 $\pm$ 0.01 & 0.27 $\pm$ 0.03 & 0.26 $\pm$ 0.02 \\
    $1^+ [1S1]$ & 0.25 & 0.28 $\pm$ 0.03 & 0.26 $\pm$ 0.02 & 0.27 $\pm$ 0.03 & 0.25 $\pm$ 0.02 \\
    \hline
  \end{tabular}
  \caption{ Mean and standard deviation of the measured $\Gamma_X$ distributions extracted from the Breit-Wigner fits. }
  \label{Table:Widths}
\end{table}

\begin{figure}
        \begin{center}
                \includegraphics[width=1.0\textwidth]{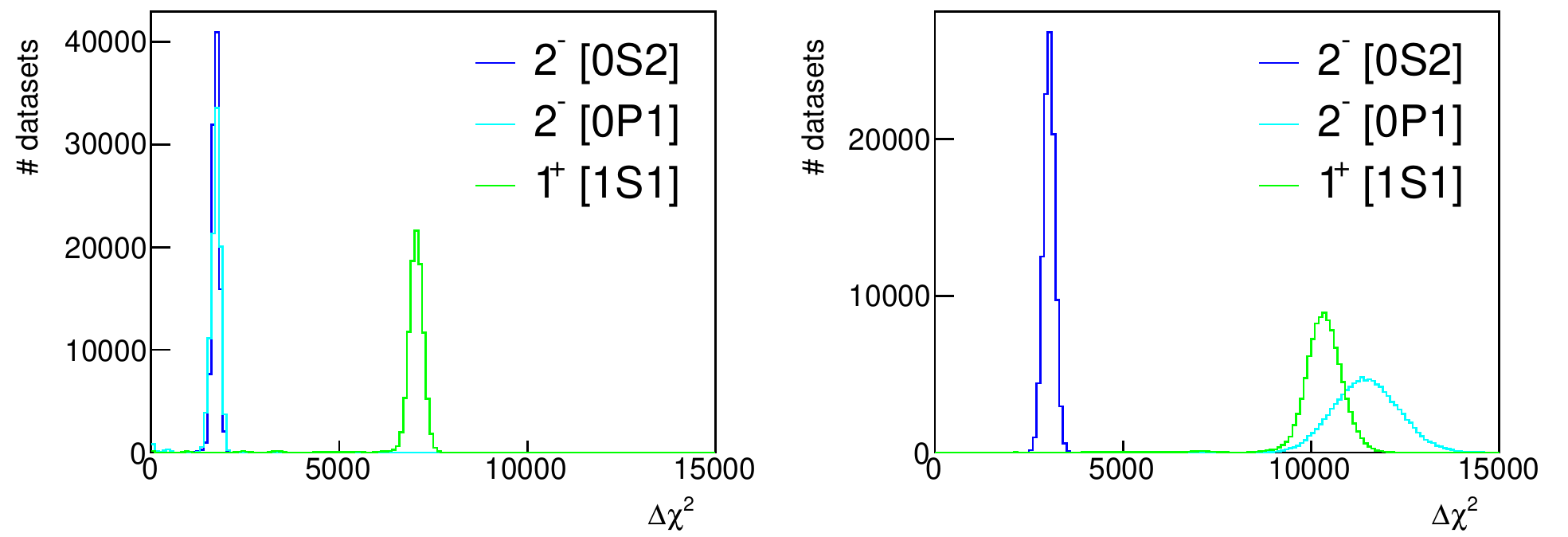}
                \caption{ $\Delta\chi^2$ distribution for resonance amplitudes in the true p.d.f.: $2^- [0S2]$ (blue), $2^- [0P1]$ (cyan), and $1^+ [1S1]$ (green); a) without using the LASSO procedure (i.e. $\lambda = 0$) and b) with the LASSO procedure using BIC to select the best $\lambda$ for each data sample and $m_{abc}$ mass bin. }
                \label{fig:deltaChi2signal}
        \end{center}
\end{figure}

We now address the topic of whether an observed resonant contribution is statistically significant.
For each resonant amplitude fit, a $\Delta\chi^2$ is calculated between the $\chi^2$ of the fit described above and the $\chi^2$ for a model with no resonant signal in the amplitude of interest.  If an amplitude did not have 6 consecutive bins with intensity $>2\sigma$ or the total phase motion, {\em i.e.}, the difference between the minimum and maximum phase difference values in the fit range defined above, was less than $\pi/4$ (another {\em ad hoc} requirement) the fit is deemed invalid and assigned $\Delta\chi^2 = 0$. 
% These choices are {\em ad hoc} and so are likely to be analysis dependent; we employ them here to have a way of identifying possible resonance contributions to be fit without the need for human inspection.
To evaluate this procedure an ensemble of $10^5$ datasets was generated.  The $\Delta\chi^2$ distribution from this ensemble for the resonant amplitudes in the true p.d.f., $2^- [0S2]$, $2^- [0P1]$, and $1^+ [1S1]$ are shown in Fig.~\ref{fig:deltaChi2signal} a) without using the LASSO procedure (i.e. $\lambda = 0$) and b) with the LASSO procedure using BIC to select the best $\lambda$ for each data sample and $m_{abc}$ mass bin.  The mean of the $\Delta\chi^2$ distribution is significantly larger for all the resonant amplitudes in the true p.d.f. when the BIC-LASSO procedure is used.  This means that a resonance will be statistically more significant when using the BIC-LASSO.

\begin{figure}
        \begin{center}
                \includegraphics[width=1.0\textwidth]{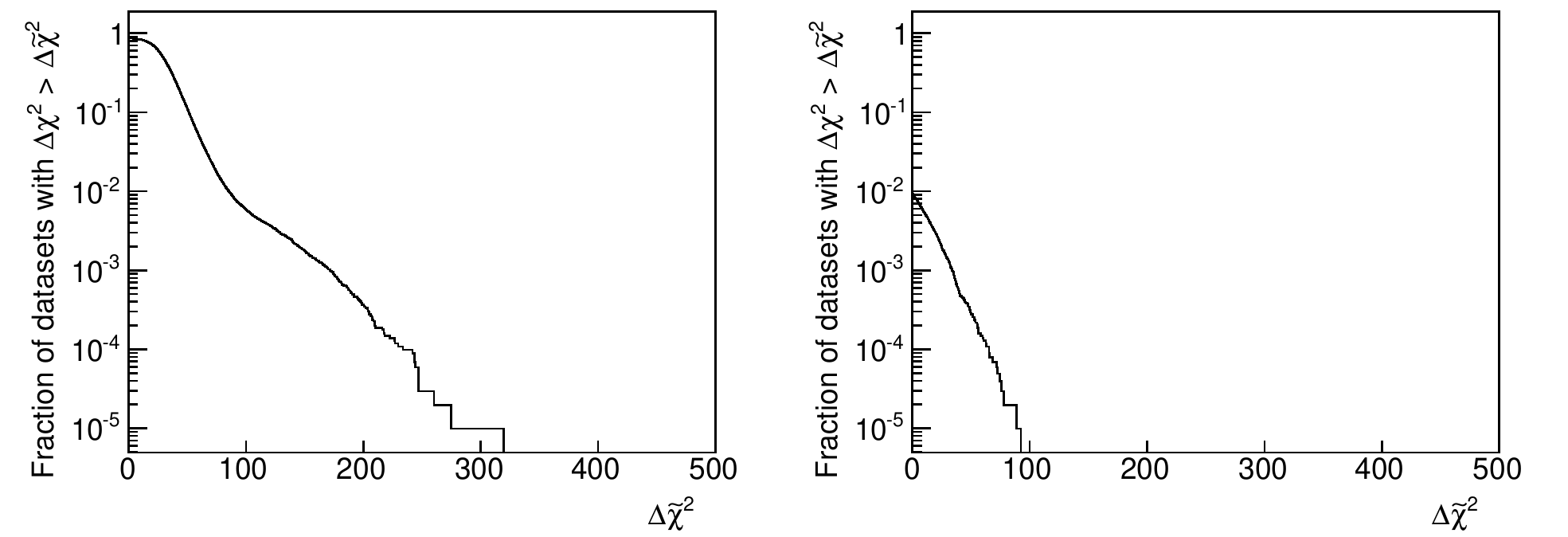}
                \caption{ Cumulative maximum $\Delta\chi^2$ distribution from the extraneous amplitudes: a) without using the LASSO procedure (i.e. $\lambda = 0$) and b) with the LASSO procedure using BIC to select the best $\lambda$ for each data sample and $m_{abc}$ mass bin. }
                \label{fig:deltaChi2extra}
        \end{center}
\end{figure}

Fig.~\ref{fig:deltaChi2extra} shows the distribution of the fraction of datasets with $\Delta\chi^2$ above a given value $\Delta\tilde{\chi}^2$ for extraneous amplitudes in the fit, a) without the LASSO procedure and b) with the BIC-LASSO procedure.  Only $\sim1\%$ of the datasets fit with the BIC-LASSO procedure had any extraneous amplitudes that satisfied the criteria for being assigned a non-zero $\Delta\chi^2$.
There are no datasets fit using the BIC-LASSO that have a maximum $\Delta\chi^2$ for any extraneous wave greater than 100.
Without using the LASSO, almost all data sets have at least one extraneous amplitude assigned a non-zero $\Delta\chi^2$ value.  Furthermore, almost 1\% have a maximum extraneous-amplitude $\Delta\chi^2 > 100$.  
In this example, $4\sigma$ significance would require at most $\Delta\chi^2 \gtrsim 250$ for the non-LASSO fits and $\gtrsim 70$ for the BIC-LASSO fits.  The actual $\Delta\chi^2$ requirements depend on the trials factor.  

%%%% Figure %%%%

%%%% Figure %%%%

In the previous section we showed that the model itself can be simplified without direct human intervention during the process which means that it is possible to get to the mass-dependent fit stage blindly.  
The type of procedure presented in this section for assigning significance could also be applied blindly, provided the procedure is fixed prior to examining the data.  
This would permit obtaining a $p$-value using pseudo data sets, since the entire procedure is automatic.  
Further research on developing a general version of such a procedure is strongly encouraged. 

%%%%%%%%%%%%%%%
%% 		Summary		 %%
%%%%%%%%%%%%%%%
\section{Summary}
\label{Sec:Summary}

This paper demonstrates how to effectively apply the LASSO regularization procedure to an amplitude analysis.  This procedure produces similar results in fits using only the true model p.d.f.\ and those that include a large number of extraneous amplitudes.
The use of regularization removes the need for direct human intervention to simplify the model, making it possible to develop procedures for determining the significance of a resonance observation.
An outline of how to assign such a significance was also presented.  
All of the techniques presented in this paper require only minor modifications to existing amplitude-analysis software.

\acknowledgments

We thank Tim Gershon for useful suggestions that improved this worked.

\clearpage

%\appendix
%\section{Additional amplitudes in the full p.d.f.}
%\label{app:fullPDF}

\end{document}